\documentclass[10pt,conference]{IEEEtran}
\usepackage[top=.71in, left=0.65in, right=0.65in, bottom=1.033in]{geometry}

\usepackage[linesnumbered,ruled,vlined]{algorithm2e}
\usepackage{paralist}
\usepackage{amssymb}

\usepackage{caption}
\usepackage{subcaption}
\usepackage{tabularx} 

\usepackage{graphicx}
\ifCLASSINFOpdf
\else
\fi
\usepackage{amsmath}
\begin{document}
%
\title{Enhanced AI as a Service at the Edge via Transformer Network}

%
%
%

\author{Vahid~Pourakbar\IEEEmembership{Student Member,~IEEE}
        and~Hamed~Shah-Mansouri~\IEEEmembership{Member,~IEEE}
		  \\ 
    Department of Electrical Engineering, Sharif University of Technology, Tehran, Iran \\
    email:  vahid.pourakbar@ee.sharif.edu, hamedsh@sharif.edu}

\maketitle

\begin{abstract}
Artificial intelligence (AI) has become a pivotal force in reshaping next generation mobile networks. Edge computing holds promise in enabling AI as a service (AIaaS) for prompt decision-making by offloading deep neural network (DNN) inference tasks to the edge. However, current methodologies exhibit limitations in efficiently offloading the tasks, leading to possible resource underutilization and waste of mobile devices’ energy. To tackle these issues, in this paper, we study AIaaS at the edge and propose an efficient offloading mechanism for renowned DNN architectures like ResNet and VGG16. We model the inference tasks as directed acyclic graphs and formulate a problem that aims to minimize the devices' energy consumption while adhering to their latency requirements and accounting for servers' capacity. To effectively solve this problem, we utilize a transformer DNN architecture. By training on historical data, we obtain a feasible and near-optimal solution to the problem. Our findings reveal that the proposed transformer model improves energy efficiency compared to established baseline schemes. Notably, when edge computing resources are limited, our model exhibits an 18\% reduction in energy consumption and significantly decreases task failure compared to existing works.

\end{abstract}

%
\IEEEpeerreviewmaketitle

\section{Introduction}
%
%
%
%
Artificial intelligence as a service (AIaaS) refers to the offering AI tools through cloud platforms, allowing the use of machine learning models easily without the hassle of creating and maintaining them. This paradigm is promising for handling intensive AI computations in emerging mobile networks and Internet of Things (IoT) systems. In addition, transitioning AI computations from centralized cloud repositories to the more proximate network edge  marks a significant paradigm shift in modern computing infrastructure. This is referred as \textit{edge AI} which can address the imminent challenges posed by remote servers latency and traffic congestion \cite{Li2020Edgent,Singh2023edgeai,Zhang2022Leung}. 


Splitting large AI models between devices and nearby servers is set to revolutionize mobile networks, making data processing faster and more efficient to meet modern digital needs. Given these advantages, utilizing immediate and localized AI services is becoming more and more common. This approach helps manage data more efficiently, reduces delay, and enhances energy-aware network designs.

Computation offloading has gained significant attention in recent years to address the increasing demand for low-latency and highly intensive deep neural network (DNN) inference tasks. Numerous studies have been conducted to assess the effectiveness of edge computing in accelerating DNN inference. These explore various aspects of computation offloading, such as task partitioning \cite{Li2023Partitioning} and resource allocation \cite{Su2023ResourceAllocation}, \cite{Fan2023ResourceAllocation}.
Xu \textit{et al.} \cite{XU2023LoadBalancing} proposed an algorithm for offloading DNN inference to the edge that aims to maximize proportional fairness through a multiple assignment problem. The proposed approach provides strong optimality guarantees with polynomial-time complexity. 
Li \textit{et al.} \cite{Li2023Inference}  focused on addressing the challenge of maximizing DNN inference throughput while accounting for delays. Their objective is to optimize the acceptance of DNN service requests by leveraging techniques such as DNN partitioning and multi-thread execution parallelism. Their research covers both offline and online request arrival scenarios, while the authors introduced an algorithm to expedite the inference process.
Gao \textit{et al.} \cite{Gao2023DNNTaskPartitioning} proposed an integrated approach to task partitioning and computational offloading for DNN inference, aiming to optimize both computation latency and energy consumption. Although this scheme demonstrates effectiveness across various DNN types, a single server is considered which restricts its applicability in distributed computing environments.

The aforementioned works exhibit limitations in efficiently allocating DNN inference tasks to appropriate computing resources. These works either overlook the DNN task partitioning or fail to properly distribute the computation among servers. These may lead to an imbalanced server load and increased delay, potentially leading to performance degradation.

In this paper, we aim to enable AIaaS on the edge while overcoming the aforementioned challenges. We consider a mobile edge computing (MEC) system where edge servers offer computing services for inference tasks of well-known DNN architectures such as ResNet18 \cite{He2015resnet} and VGG16 \cite{Simonyan2014vgg16}. We model these DNN inference tasks as directed acyclic graphs (DAGs) and propose an algorithm to effectively schedule and offload them. 
We introduce a novel approach that employs a transformer DNN architecture to leverage its power in solving large-scale combinatorial optimization problems as inspired by \cite{kool2019attention}, \cite{Vaswani2017Attention}, \cite{Bahrami2022Wong}. Compared to existing algorithms, our approach, by learning heuristic rules from training data, shows improved capability in obtaining optimal solutions for the offloading problem in the heterogeneous MEC environment.

The main contributions of this work are as follows.
\begin{itemize}
\item \textit{Problem formulation:} We formulate an optimization problem that aims to minimize the total energy consumption associated with DNN inference tasks, while ensuring timely task completion. Our approach effectively tackles the heterogeneity of DNN layers and their unique computational characteristics. 

\item \textit{DNN transformer:} To efficiently solve the formulated combinatorial optimization problem, we propose a novel transformer architecture. We design a transformer to handle the inherent combinatorial nature of the problem, allowing it to capture intricate dependencies and relationships within the DAGs representing DNN inference tasks. By training the transformer with a sufficient number of sample problem instances using historical data, we achieve the capability to derive a feasible and near-optimal solution for new instances of the problem within polynomial time. This approach not only ensures the efficiency of our algorithm, but also enhances its effectiveness in providing accurate and efficient solutions. 
\item \textit{Performance evaluation:} We thoroughly assess the effectiveness of our proposed algorithm. The results demonstrate its efficiency in saving devices' energy while maintaining a lower task failure rate. Specifically, when computing resources at the edge are limited, our model achieves a significant reduction in energy consumption compared to baseline schemes. It also reduces the energy consumption per completed task by 18\% compared to existing works \cite{Gao2023DNNTaskPartitioning}, \cite{Zhang2023datasetpower}, \cite{Liu2022loadpart}. This attests to the superior energy efficiency of our algorithm and underscores its potential in reshaping AIaaS at the edge.
\end{itemize} 

This paper is organized as follows. Section \ref{section:systemmodel} presents the system model and problem formulation. The methodology is proposed in Section \ref{section:methodology}, which encompasses the transformer DNN architecture and dataset generation. Simulation results are provided in Section \ref{section:performance} and Section \ref{section:conclusion} concludes the paper.


\section{System Model and Problem Formulation} \label{section:systemmodel}

\subsection{System Overview}
\begin{figure}[t]
\centering
  \includegraphics[width=.65\linewidth]{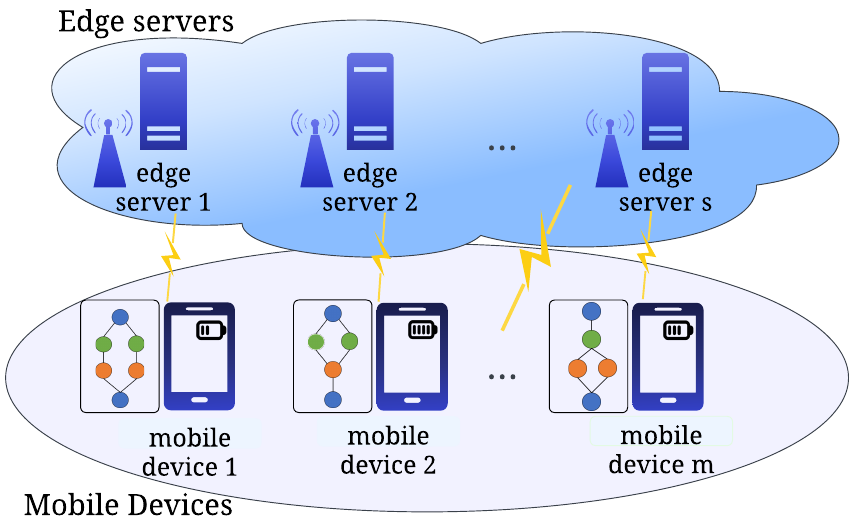}
  \caption{An instance of the MEC system with edge AI.}
  \label{fig:mobile_server}
  \vspace{-4mm}
\end{figure}
\begin{figure}[t]
\centering
  \includegraphics[width=0.65\linewidth]{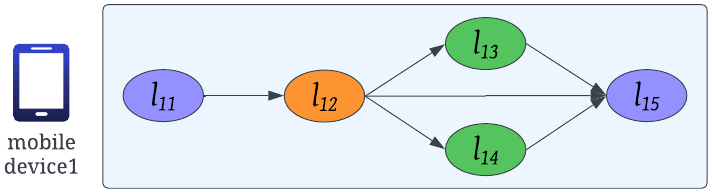}
  \caption{DNN task of a mobile device, showcasing the layer dependency.}
  \label{fig:sample_layers}
  \vspace{-4mm}
\end{figure}

We consider a pool of edge servers providing computing resources to mobile devices for DNN inference. The tasks can be either  executed locally on the device or offloaded to the edge servers. Fig.~\ref{fig:mobile_server} illustrates the system where we consider a set of $M$ mobile devices, denoted as $\mathcal{M}=\{1,2,...,M\}$. Each mobile device $m \in \mathcal{M}$ has a DNN inference task
where each task
is represented as a DAG $\mathcal{G}_m = (\mathcal{L}_m, \mathcal{E}_m)$, where $\mathcal{L}_m$ and $\mathcal{E}_m$ represent the graph nodes and edges, respectively. 
The nodes of device $m$'s DAG $\mathcal{G}_m$ represent the layers of the corresponding DNN. Moreover, we define the set $\mathcal{L}_m = \{l_{m,1}, l_{m,2}, \ldots, l_{m,L}\}$, where $l_{m,v}$ denotes the $v$th layer of DNN task while $L$ is the number of such layers. The edges in the DAG, defined as $(l_{m,i}, l_{m,j})$, indicate the dependencies between the layers. Specifically, if there is an edge from layer $l_{m,i}$ to $l_{m,j}$, it implies that $l_{m,j}$ depends on $l_{m,i}$ and must be evaluated before proceeding to the next layer as shown in Fig.~\ref{fig:sample_layers}. 
The set of edge servers is denoted as $\mathcal{S} = \{1, 2, ..., S\}$, where $S$ is the total number of edge servers available. Additionally, each edge server is characterized by its processing power, denoted as $C_s$ in terms of floating point operation per second (FLOPS).
The mobile devices and edge servers together form the set of all computing resources which we denote as $\mathcal{U} \triangleq \mathcal{M} \cup \mathcal{S}$. Each computing resource $u \in \mathcal{U}$ is characterized by its computing resource capacity, denoted as $q_u$. The value of $q_u$ quantifies the processing capability of the computing resources in terms of CPU cycles per second.

To account for the transmission rate between mobile devices and edge servers, as well as between edge servers, we introduce $b_{k,j}$, which is the transmission rate between the mobile device or server $k \in \mathcal{U}$ and server $j \in \mathcal{S}$. We further define matrix $\mathbf{B} = (b_{k,j})_{k \in \mathcal{U}, j \in \mathcal{S}}$
and denote the $v$th layer of device $m$'s DNN as tuple $l_{m,v} = \langle a_{m,v}, I_{m,v}, O_{m,v} \rangle$, where $a_{m,v}$ is the number of floating-point operation for $l_{m,v}$ \cite{Chen2022SystemModel}. This characterizes the layer's computational workload, while $I_{m,v}$ and $O_{m,v}$ respectively indicate the input size and output size of the layer.
For an individual convolutional layer $v$ performed on device $m$, the required number of floating-point operations denoted by $F^{\text{conv}}_{m,v}$ is \cite{XU2023LoadBalancing}:
\begin{align}
F^{\text{conv}}_{m,v} = 2 I^{h}_{m,v} I^{w}_{m,v} (c^{\text{in}} K^{w}_{m,v} K^{h}_{m,v} + 1) c^{\text{out}}. \label{eq:F_conv}
\end{align}

In (\ref{eq:F_conv}), $I^{h}_{m,v}$ and $I^{w}_{m,v}$ indicate the height and width of the input feature map, correspondingly, whereas $K^{w}_{m,v}$ and $K^{h}_{m,v}$ stand for the width and height of the convolutional layer $v$'s kernel. Additionally, $c^{\text{in}}$ represents the number of channels within the input feature maps, while $c^{\text{out}}$ signifies the number of channels in the output feature maps.

In the case of fully connected layers, the total floating-point computations for layer $v$ on device $m$ is given by $F^{\text{fc}}_{m,v}$ \cite{XU2023LoadBalancing}.
\begin{align}
F^{\text{fc}}_{m,v} = I_{m,v} O_{m,v} + O_{m,v} (I_{m,v} - 1). 
\end{align}
It is worth noting that for layer $v$, the calculated amount $a_{m,v}$ is equal to $F^{\text{conv}}_{m,v}$ when a convolutional layer is present, and it is equal to $F^{\text{fc}}_{m,v}$ if a fully connected layer is used. However, $a_{m,v}$ is generally negligible compared to convolutions and dot products, so its impact on the total number of operations can be disregarded similar to \cite{XU2023LoadBalancing}.

The execution time of device $m$'s layer $v$ on computing resource $u$, denoted as $T^{\text{exe}}_{m,v,u}$, is calculated as follows \cite{Chen2022SystemModel}.
\begin{align}
T^{\text{exe}}_{m,v,u} = \frac{a_{m,v}}{q_u}.
\end{align}
For the case where $u \in \mathcal{M}$, we introduce $T^{\text{exe, local}}_{m,v,u}=T^{\text{exe}}_{m,v,u}$. Similarly, for $u \in \mathcal{S}$, we define $T^{\text{exe, server}}_{m,v,u}=T^{\text{exe}}_{m,v,u}$.
Let $T^{\text{trans}}_{m,v,u}$ denote the time required for transmitting layer $v$ to server $u$ and obtaining the resulting output. We have
\begin{align}
T^{\text{trans}}_{m,v,u} = \frac{I_{m,v} + O_{m,v}}{b_{m,u}}.
\end{align}

The energy consumption for executing and transmitting data of layer $v$ on device $m$ using computing resource $u$ can be expressed using the following equations.
\begin{align}
E_{m,v,u}^{\text{exe}} = P^{\text{exe}}_{u} T^{\text{exe}}_{m,v,u}, \\
E_{m,v,u}^{\text{trans}} = P^{\text{trans}}_{m, u} T^{\text{trans}}_{m,v,u}.
\end{align}
Here, $P^{\text{exe}}_{u}$ represents the power consumption for executing the layer on computing resource $u$, and $P^{\text{trans}}_{m, u}$ denotes the power used for data transmission between mobile device $m$ and server $u$ during the execution of layer $v$.

\subsection{Problem Formulation}
This paper explores optimizing energy consumption on mobile devices when running DNN tasks with limited resources, addressing the challenge of insufficient computational capabilities leading to prolonged execution times and high energy consumption. It investigates offloading certain DNN layers to edge servers with sufficient computational resources as a potential solution to reduce energy consumption, while also considering local execution on end devices if feasible.
To make decisions regarding layer execution, we introduce a binary decision variable $x_{m,v,u}$, which takes a value of 1 if layer $v$ is offloaded from device $m$ to server $u$, and 0 otherwise. Additionally, we introduce another binary decision variable \(z_{m,v,u}\), where \(z_{m,v,u}\) equals 1 when layer  $v$ of device $m$ is uploaded to or downloaded from edge server $u$, and 0 when the layer continues its execution on the same device. Note that \(z_{m,v,u}\) is constrained as follows.
\begin{equation}
z_{m,v,u} \geq |x_{m,v,u} - x_{m,v-1,u}|, \forall m \in \mathcal{M}, v \in \mathcal{L}_m, u \in \mathcal{U}. \label{eq:transfer_binery}
\end{equation}
To calculate the total execution time for the DNN task, we sum up the execution times for all layers and the time to transfer data between mobile device and edge server.
\begin{align}
T^{\text{total}}_{m} &= \sum_{v \in \mathcal{L}_m} \sum_{u \in \mathcal{U}} \left((1 - x_{m,v,u}) T^{\text{exe, local}}_{m,v,u} \right. \nonumber \\
&\quad+ z_{m,v,u} T^{\text{trans}}_{m,v,u} + x_{m,v,u}T^{\text{exe, server}}_{m,v,u}\big). \label{eq:task_time}
\end{align}

The total energy consumption of layer $v$ of device $m$ can be calculated by summing up the energy spent during the execution of the layer and the energy consumption for data transmission between device and edge servers. We have
\begin{align}
E_{m,v} = \sum_{u \in \mathcal{U}} \left((1 - x_{m,v,u}) E_{m,v,u}^{\text{exe}} + z_{m,v,u} E_{m,v,u}^{\text{trans}}\right).
\end{align}

We need to consider the following constraints.

1. Task deadline: The offloading decision must ensure that the total task execution time, encompassing both computation at the computing resources and data transfer, is shorter than the task's deadline denoted by $\tau_{m}$. Offloading a layer $v$ is only considered advantageous if the total task completion time meets the task deadline $\tau_{m}$.

2. Edge server capacity: Each edge server has a limited capacity for executing layers. We incorporate a constraint to ensure that the total number of layers offloaded to an edge server does not exceed its capacity.


Our objective is to minimize the total energy consumption of executing the DNN on mobile devices. We now express the optimization problem as follows.
\begin{subequations} \label{eq:OurProblem}
\begin{flalign}
\text{minimize}  & \sum_{m, v, u} (1 - x_{m,v,u}) E_{m,v,u}^{\text{exe}} + z_{m,v,u} E_{m,v,u}^{\text{trans}} & \label{eq:objective} \\
\text{subject to } & \text{constraints } (\ref{eq:transfer_binery})\text{-}(\ref{eq:task_time}), & \label{eq:constraintRef} \\
& T^{\text{total}}_{m} \leq \tau_{m}, \; \forall m \in \mathcal{M}, & \label{eq:taskdeadline} \\
& \sum_{m, v} x_{m,v,u} a_{m,v} \leq C_s, \; \forall s \in \mathcal{S}, & \label{eq:max_capacity} \\
& \sum_{u \in \mathcal{U}} x_{m,v,u} \leq 1, \; \forall m \in \mathcal{M}, v \in \mathcal{L}_m, & \label{eq:avoidduplication} \\
& x_{m,v,u}, z_{m,v,u} \in \{0, 1\}, \; \forall m \in \mathcal{M}, v \in \mathcal{L}_m, u \in \mathcal{U}. & \label{eq:binarydecision}
\end{flalign}
\end{subequations}

Here, \eqref{eq:taskdeadline} ensures that for each device $m$, the overall execution time, including computation and data transfer of all layers meets the deadline $\tau_{m}$. Constraint \eqref{eq:max_capacity} restricts the number of offloaded layers to each edge server $u$ so that it does not surpass its capacity $C_s$, preventing overload. Moreover, constraint \eqref{eq:avoidduplication} ensures that a layer $v$ of mobile device $m$'s task is assigned to only one edge server $u$ at a time.

Problem (\ref{eq:OurProblem}) is a combinatorial problem, which is inherently difficult to solve. Therefore, we propose an algorithm to learn the best policies and achieve a near-optimal solution.

\section{Proposed Transformer Architecture} \label{section:methodology}
To tackle the inherent combinatorial nature of the formulated problem, we propose a transformer-based DNN architecture.
The DNN transformers are mostly used for language processing and  translation. By considering the mobile inference tasks as sentences and layers as words within the sentence, we can leverage the advantages of the transformer to solve the problem, inspired by the effectiveness of transformers \cite{kool2019attention} \cite{Vaswani2017Attention}. 
Thus, we develop our proposed transformer-based DNN to capture complex dependencies and relationships within the DAGs that represent different DNN inference tasks. By training the transformer DNN using historical data from sample problem instances, we equip it with the ability to efficiently solve new instances of the optimization problem.
This helps learn rules that adjust the offloading decisions for DNN layers. The learned rules highly contribute to finding optimal solutions in the dynamic and uncertain MEC environment, ultimately resulting in optimal energy consumption.

\subsection{Transformer DNN Architecture Overview}
\begin{figure}[t]
\centering
  \includegraphics[width=0.85\linewidth]{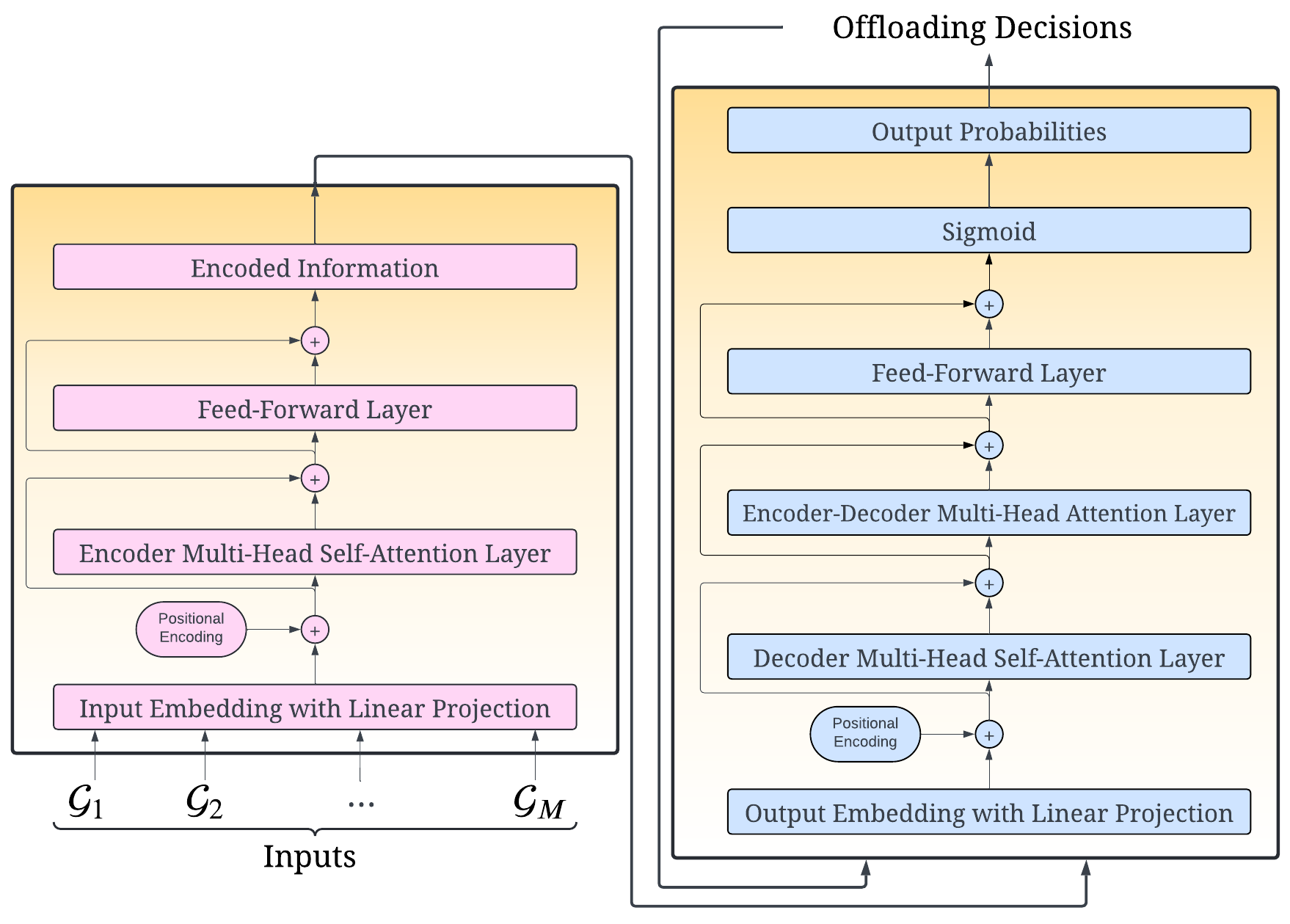}
  \caption{Proposed transformer architecture for task offloading problem.}
  \label{fig:transformer_architecture}
\end{figure}

We employ a transformer-based DNN architecture, as shown in Fig.~\ref{fig:transformer_architecture}, specifically adapted to address the complexities of processing information related to DNN inference layers. This includes managing the computational requirements and offloading costs associated with each layer. Encoders extract sequential layer information from the task DAGs, while decoders are responsible for making offloading decisions.
During the training phase, the input consists of $M$ DAGs representing task information. This input is passed through an embedding layer using linear projection, which also integrates positional encoding as described by \cite{kool2019attention}. The encoder architecture includes a multi-head self-attention layer followed by a feedforward network. 
The decoder receives the target sequences along with the encoder's output. It comprises multiple multi-head attention layers and feedforward networks. The primary attention mechanism within the decoder focuses on the encoder's output to ensure contextual relevance in offloading decisions. Subsequent layers in the decoder refine the output sequence by incorporating information from previous decisions, ensuring that each offloading decision is coherent and contextually appropriate.
At the decoder's output, a sigmoid activation function is employed to generate binary decisions for each element within the $L$-element sequence, indicating a 0 for local execution or a 1 for offloading to edge servers. To mitigate overfitting, we apply regularization techniques such as dropout and use a predetermined teacher-forcing ratio during training, where the decoder's target input is occasionally replaced with the actual target output. During inference, autoregressive decoding is performed within the decoder \cite{wu2023teacherforcing}. The resulting output sequence consists of $M$ elements, each with $L$ binary decisions, indicating the task layers designated for offloading.

\subsection{Dataset Generation and Algorithm Design}
For generating dataset, we focus on the following three pivotal image processing tasks: MobileNetV1 \cite{Howard2017mobilenet}, ResNet18, and VGG16. We provide the information of these tasks on mobile devices, arrays with a length equal to the number of task layers (i.e., $L$), containing information about each layer, as input to our transformer. The offloading decisions for each layer of tasks are then determined.

\begin{algorithm}[ht]
    \small
    \DontPrintSemicolon
    \caption{\small Transformer-Based Learning Algorithm for DNN Task Offloading}
    \label{algorithm:transformer_model}
    \KwIn{Inference tasks $\mathcal{G}_m$, $m \in \mathcal{M}$, Number of training epochs $I$, Number of batches per epoch $B$, Training epoch size $E^{\text{train}}$, Evaluation size $E^{\text{eval}}$}
    Initialize epoch index $i := 1$, Set step index $j := 1$\;
    Initialize neural network parameter $\theta_1$;

    \While{$i \leq I$}{
        Select set $D_i^{\text{train}} \subseteq D^{\text{train}}$ of $E^{\text{train}}$ sample data randomly\;
        Select set $D_i^{\text{eval}} \subseteq D^{\text{eval}}$ of $E^{\text{eval}}$ sample data randomly\;
        Divide set $D_i^{\text{train}}$ into $K$ batches $D_{i, k}^{\text{train}}$ for $k \in \mathcal{K}$\;
        Set batch index $k := 1$\;

        \While{$k \leq K$}{
            Give inference task layer info of mobile devices to the transformer model for sample data $d \in D_{i, k}^{\text{train}}$\;
            Obtain output sequence $x_{m,v,u}$ as the offloading decision for each layer of the task for sample data $d \in D_{i, k}^{\text{train}}$\;
            Compute the loss function using the penalized form of problem (\ref{eq:OurProblem}) for sample data $d \in D_{i, k}^{\text{train}}$\;
            Update neural network parameter $\theta_j$ in step $j$ using the Adam optimizer\;
            Update batch and step index $k := k + 1$, $j := j + 1$\;
        }

        Evaluate the updated model on the evaluation set $D_{i}^{\text{eval}}$\;
        Update epoch index $i := i + 1$\;
    }
    \KwOut{Offloading decisions}
\end{algorithm}

Algorithm \ref{algorithm:transformer_model} shows our proposed DNN transformer model. We initialize the number of training epochs $I$, the number of batches per epoch $K$, and denote the set of batch indices as $\mathcal{K}=\{1,2,...,K\}$. Additionally, we define the training epoch size $E^{train}$ and the evaluation size $E^{eval}$. We use the sets $D^{train}$ and $D^{eval}$ of sample problem instances for training and evaluating the proposed DNN, respectively. Each sample problem instance comprises historical data for the task graphs $\mathcal{G}_m$, which include information about computation workload, layer type, and size. Lines 1 and 2 of the algorithm handle initialization and introduce the training attributes. During each of the specified training epochs (Line 3), the model randomly selects training and evaluation data sets, and divides the training data into batches (Lines 4-7). The core operation in each batch involves providing inference task layer information to the transformer model and obtaining an offloading decision sequence, which is further refined by computing a loss function and updating the neural network parameters using the Adam optimizer (Lines 8-15). Upon processing all batches for an epoch, the model's performance is evaluated using the evaluation subset. The algorithm concludes by providing an optimized offloading decision for each task of mobile devices.

\section{Performance Evaluation} \label{section:performance}
In this section, we assess the performance of our proposed DNN transformer algorithm and compare it with partitioning algorithms proposed in \cite{Gao2023DNNTaskPartitioning}, \cite{Zhang2023datasetpower}, \cite{Liu2022loadpart}. This algorithm identifies a partitioning point within the layers of a DNN task and offloads only the layers beyond that point. For a comprehensive understanding, we also compare against local execution and offloading only baseline schemes, as well as a random algorithm that arbitrarily selects layers for offloading.

Unless stated otherwise, we consider 30 mobile devices with CPU frequencies uniformly chosen from the range [0.5, 1] GHz. We also consider 9 edge servers with processing powers uniformly distributed within [1.5, 3] GHz similar to \cite{tian2023datasetcpu}. The constant power consumption of mobile devices, denoted as \(P^{\text{exe}}_{m}\), uniformly varies between [2, 6] Watts, whereas the transmission power, \(P^{\text{trans}}_{m}\), is defined as \(1.01 P^{\text{exe}}_{u}\) \cite{Zhang2023datasetpower}. The servers' capacity, represented as \(C_s\), ranges from 4 to 32 GFLOPS. Task deadlines, ranging from 1 to 5 seconds, are customized based on the task's complexity, with more complex tasks like VGG16 allocated longer duration compared to simpler ones like MobileNetV1. 

\begin{figure}[t]
  \includegraphics[width=0.85\linewidth]{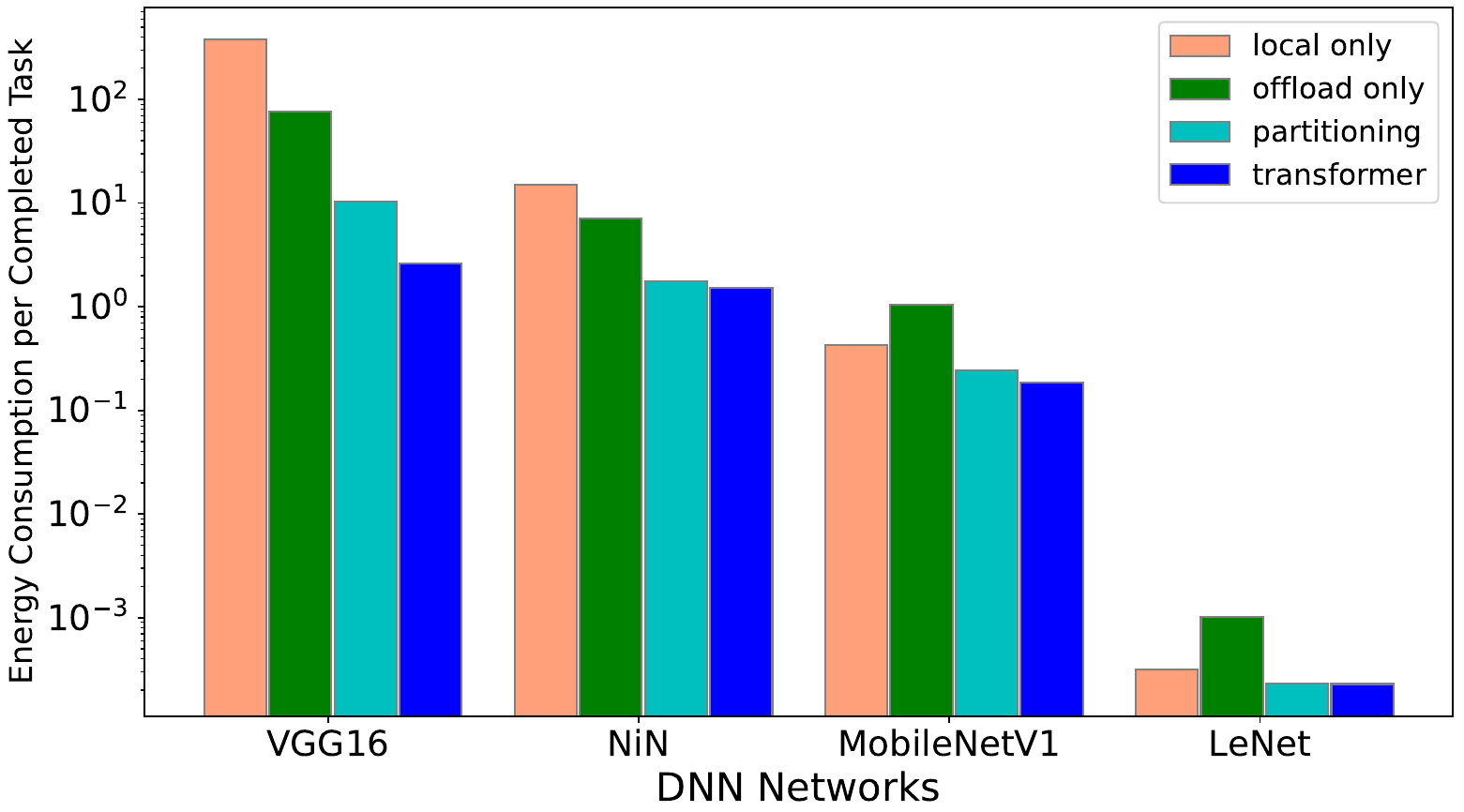}
  \centering
  \caption{Average energy consumption for completing a task for different algorithms under various DNN models. To ensure a fair comparison, the energy consumed by failed tasks is excluded.}
  \label{fig:task_type}
  \vspace{-6mm}
\end{figure}

\begin{figure*}[t]
  \includegraphics[width=\linewidth]{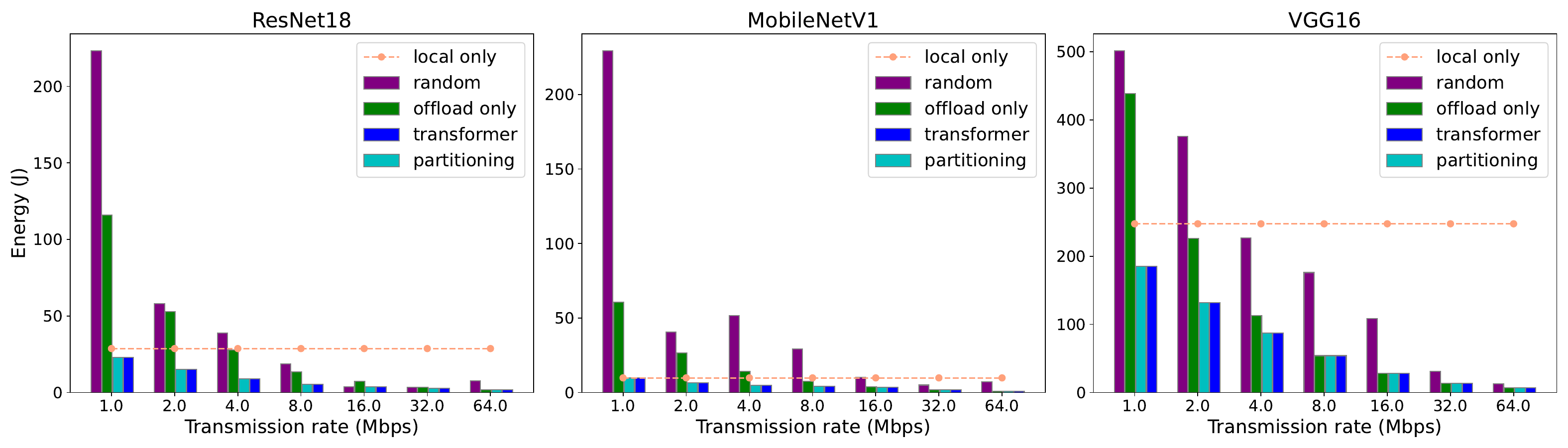}
  \caption{Energy consumption across different transmission rates for various DNN tasks:  a) ResNet18, b) MobileNetV1, c) VGG16.}
  \label{fig:3plot_enough_resources}
\vspace{-4mm}
\end{figure*}


To test our proposed model's adaptability across diverse task types and conditions, we consider additional image processing tasks, namely NiN \cite{Lin2014nin} and LeNet \cite{Lecun1998lenet}, beyond those initially trained for mobile devices.
Fig.~\ref{fig:task_type} shows the total energy consumption of mobile devices, each executing the same DNN task, normalized by the number of completed tasks. We compare the following algorithms: 
\begin{inparaenum}[a)]
\item local only task execution,
\item offloading only,
\item partitioning,
\item and our proposed transformer-based algorithm.
\end{inparaenum} 
The figure reveals an inverse correlation between a DNN's computational complexity and its energy consumption. Specifically, complex DNNs like VGG16 require substantial computational resources when executed locally on mobile devices, which have limited processing capabilities. Consequently, offloading these tasks to edge servers is more energy-efficient. Conversely, less complex tasks like LeNet may become inefficient and energy-intensive when offloaded \cite{Zhang2023datasetpower}. Notably, our proposed transformer-based algorithm consistently outperforms the other three algorithms for all evaluated DNN models, achieving significant reductions in energy consumption.

We now evaluate the effects of different data transmission rates on the performance of DNN inference tasks. Low transmission rates emphasize scenarios that favor local computing, whereas higher transmission rates encourage devices to offload their tasks. We consider a  mobile device executing DNN inference tasks, namely ResNet18, MobileNetV1, and VGG16, interacting with an edge server. 
Fig.~\ref{fig:3plot_enough_resources} illustrates the energy consumption of our proposed transformer-based algorithm in comparison with the other four algorithms. We can observe that for ResNet18, the model prefers local execution due to the inefficiency of offloading in low transmission rates. With increasing transmission rate, the model begins offloading layers to edge servers. At high transmission rates, the algorithm offloads all layers. Additionally, it can be observed that the energy consumption for our transformer model consistently outperforms random offloading and local only execution across all transmission rates.
The behaviors observed for MobileNetV1 and VGG16 are consistent with the ResNet18. Our model performs similarly to the partitioning algorithm when there are sufficient computational resources. It determines the optimal partition point and offloads subsequent layers to reduce energy consumption, all while adhering to task deadlines.

\begin{figure}[t]
    \centering
    \begin{subfigure}[b]{0.8\linewidth}
        \includegraphics[width=\linewidth]{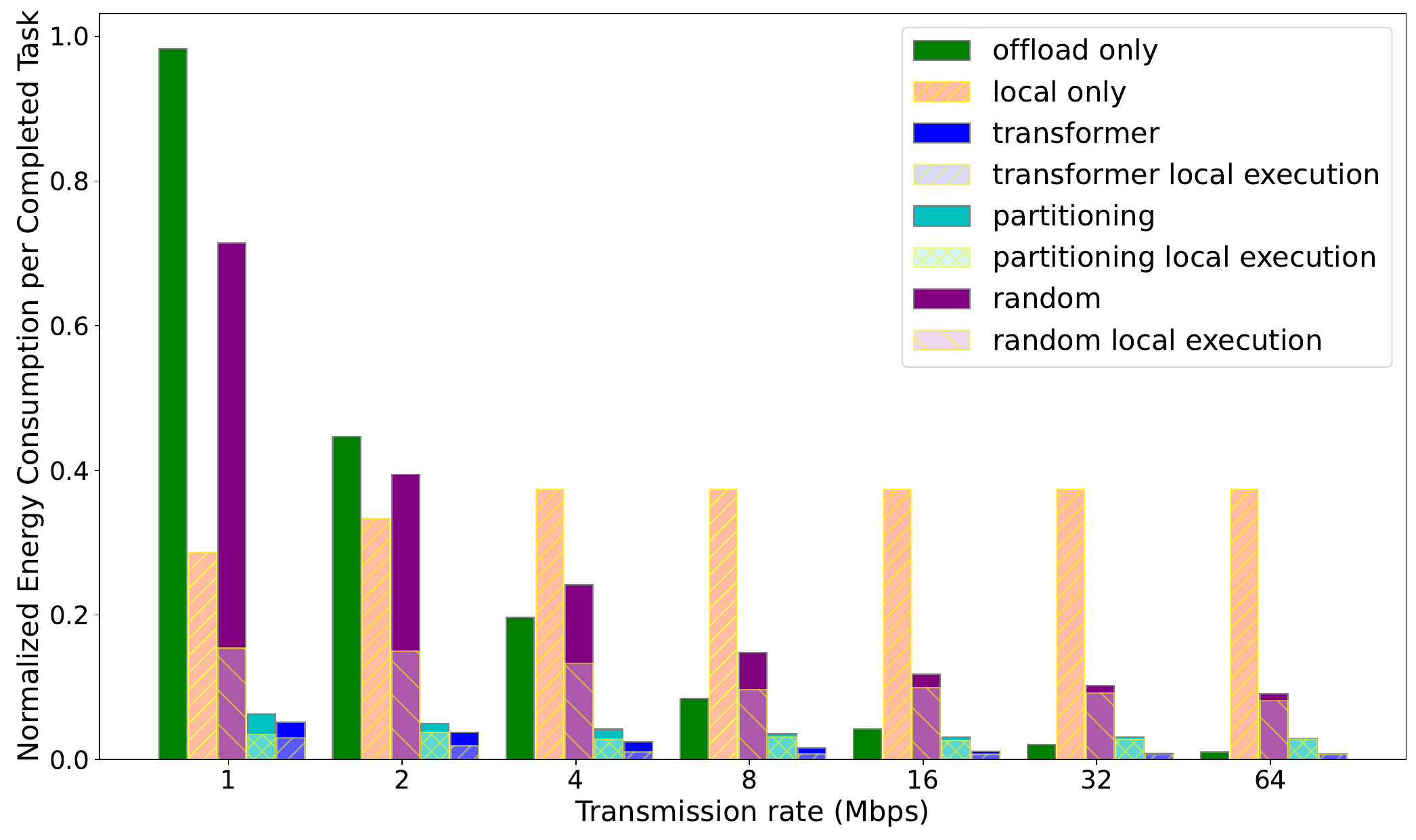}
        \caption{}
        \label{fig:1_energy_bandwidth4}
    \end{subfigure}
    
  
    \begin{subfigure}[b]{0.9\linewidth}
        \includegraphics[width=\linewidth]{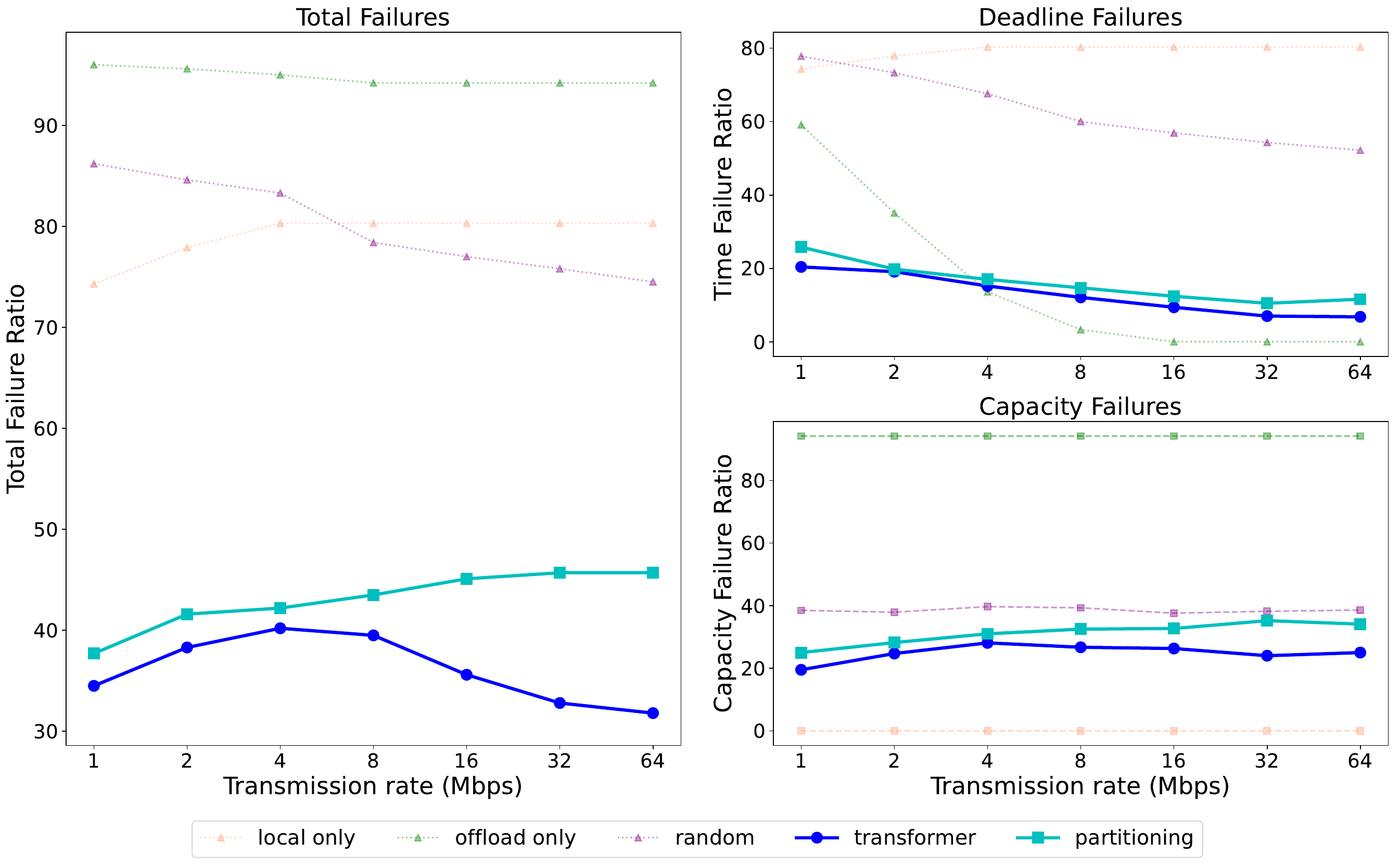}
        \caption{}
        \label{fig:fails_bandwidth_1}
    \end{subfigure}

    \caption{(a) Energy consumption per completed task, (b) Task failure rates under varying transmission rates.}
    \label{fig:performance_comparison}
    \vspace{-6mm}
\end{figure}

Challenges arise in scenarios with limited computational resources. To illuminate our findings, the normalized energy consumption per completed task across diverse transmission rates for five distinct scenarios is depicted in Fig.~\ref{fig:1_energy_bandwidth4}. From this figure, it is evident that higher transmission rates lead to lower energy consumption across all algorithms except for the local only scenario. Interestingly, the decline in the energy consumption for our model is more significant. Our model demonstrates a reduction in normalized energy consumption ranging from 18\% at a 1 Mbps transmission rate to approximately 73\% at 64 Mbps compared to the partitioning algorithm. Fig.~\ref{fig:fails_bandwidth_1} shows the percentage of the tasks failed due to violating the constraints of problem (\ref{eq:OurProblem}).
The results indicate that as the transmission rate increases, the deadline failure ratio declines. This trend suggests that the model offloads tasks more frequently, leading to a subsequent rise in failures due to limited edge server capacity. Furthermore, our transformer model consistently outperforms the partitioning algorithm, resulting in fewer failures in scenarios with both capacity constraints and time deadlines. This comparative performance underscores the inherent limitations of the partitioning algorithm, particularly when deployed in environments with constrained edge servers or an overwhelming number of tasks. These cause even more challenges at higher transmission rates.
Note that although the offloading only algorithm reduces the energy consumption when transmission rate is high, it actually results in more task failure due to limited capacity.

\section{Conclusion} \label{section:conclusion}
In this paper, we proposed an offloading algorithm for DNN inference tasks in MEC to facilitate AIaaS. We first formulated a combinatorial optimization problem to minimize energy while satisfying latency constraints. We then introduced a task offloading mechanism leveraging a transformer DNN architecture trained on historical data. Although the transformer model matched baseline schemes in environments with sufficient computing resources, it showed substantial energy savings compared to standard benchmarks in resource-constrained settings. Furthermore, it outperforms an existing algorithm by reducing energy consumption per completed task by 18\% and decreasing task failure rates when computing resources at the edge are limited. These interesting findings make our proposed algorithm a good choice for deploying AIaaS in the edge.



%





\ifCLASSOPTIONcaptionsoff
  \newpage
\fi



%

\bibliographystyle{IEEEtran}
\bibliography{ref.bib}

%








\end{document}